# Optimized Graphene Electrodes for contacting Graphene Nanoribbons

Oliver Braun,[a,b] Jan Overbeck,[a,b,c] Maria El Abbassi,[a,b,d] Silvan Käser,[a,e] Roman Furrer,[a] Antonis Olziersky,[f] Alexander Flasby,[a] Gabriela Borin Barin,[g] Rimah Darawish,[g,h] Klaus Müllen,[i] Pascal Ruffieux,[g] Roman Fasel,[g,h] Ivan Shorubalko,[a] Mickael L. Perrin,[a,*] and Michel Calame,[a,b,c,**]

- a. Transport at Nanoscale Interfaces Laboratory, Empa, Swiss Federal Laboratories for Materials Science and Technology, 8600 Dübendorf, Switzerland
- b. Department of Physics, University of Basel, 4056 Basel, Switzerland
- c. Swiss Nanoscience Institute, University of Basel, 4056 Basel, Switzerland
- d. Kavli Institute of Nanoscience, Delft University of Technology, 2628 CJ Delft, The Netherlands
- e. Department of Chemistry, University of Basel, 4056 Basel, Switzerland
- f. IBM Research – Zurich, 8803 Rüschlikon, Switzerland
- g. nanotech@surfaces Laboratory, Empa, Swiss Federal Laboratories for Materials Science and Technology, 8600 Dübendorf, Switzerland
- h. Department of Chemistry and Biochemistry, University of Bern, 3012 Bern, Switzerland
- i. Max Planck Institute for Polymer Research, 55128 Mainz, Germany

Corresponding authors:

*E-Mail: mickael.perrin@empa.ch

**E-Mail: michel.calame@empa.ch

Keywords:

graphene nanoribbons, graphene electrodes, thermal annealing, Raman spectroscopy, field-effect transistor

**Atomically precise graphene nanoribbons are a promising emerging class of designer quantum materials with electronic properties that are tunable by chemical design. However, many challenges remain in the device integration of these materials, especially regarding contacting strategies. We report on the device integration of uniaxially aligned and non-aligned 9-atom wide armchair graphene nanoribbons (9-AGNRs) in a field-effect transistor geometry using electron beam lithography-defined graphene electrodes. This approach yields controlled electrode geometries and enables higher fabrication throughput compared to previous approaches using an electrical breakdown technique. Thermal annealing is found to be a crucial step for successful device operation resulting in electronic transport characteristics showing a strong gate dependence. Raman spectroscopy confirms the integrity of the graphene electrodes after patterning and of the GNRs after device integration. Our results demonstrate the importance of the GNR-graphene electrode interface and pave the way for GNR device integration with structurally well-defined electrodes.**

# 1 Introduction

New classes of electronic nanomaterials often require several years to decades of research to develop reliable electrical contacting approaches. For example, it took more than two decades to go from the first carbon nanotube (CNT) field-effect transistors to their successful integration into microprocessors.[1–5] Similar timescales were also needed to develop the field of semiconducting nanowires from the first reporting of Si-whiskers to their reliable use for quantum computing.[6–8] In the case of Si-nanowires, surface passivation of the contact area and thermal annealing were found to increase device performance significantly.[9] More recently, bottom-up synthesized atomically precise graphene nanoribbons (GNRs) have attracted a lot of attention as their electronic and magnetic properties can be tailored by bottom-up synthesis. However, contacting GNRs using top town fabrication processes turned out to be highly demanding, in particular, because of their nanoscale dimensions of around 1 nm in width and lengths reaching rypicall 5-50 nm.[10–15]

Standard electron beam lithography and metallization processes have been used by several groups for contacting GNRs.[12, 16, 17] However, these methods involve processing and metallization on top of transferred GNRs and can lead to the introduction of contaminants at the contact-GNR interface and/or GNR damage. This approach is particularly problematic for GNRs with reactive and/or functionalized edges.[18–21] Alternatively, GNRs have also been transferred on top of predefined metal electrodes.[22] This approach may be suitable for GNR films in which hopping of charge carriers over larger distances is the dominating effect on transport properties but may lead to ill-defined 3-dimensional junction geometries when contacting a single GNR. Furthermore, metal electrodes in short-channel devices lead to the formation of image charges and screen the applied electrostatic gate field used to tune the electronic transport, requiring advanced gating approaches such as ionic liquid gating for reaching a sufficient gating efficiency.[12, 23–25] Finally, we expect that the disorder of metallic electrodes at the atomic scale leads to uncontrolled local electrostatic potential surrounding the nanoscale object, a problem that 2D covalent crystals have the potential to overcome.

The above-mentioned issues can be adressed by the use of graphene electrodes. Graphene, with its monoatomic thickness, allows for the GNRs to be transferred on top of the electrodes, without introducing significant bending of the GNRs bridging the source and drain electrodes. The π-π orbital overlap is widely used for contacting two-dimensional materials [26, 27] and the charge carrier density in the graphene leads can be tuned by electrostatic gating. Graphene electrodes fabricated using the well-established electrical breakdown procedure result in gaps separating the electrodes by a few nanometers and are a suitable way to contact graphene nanoribbons of various types.[11, 13, 28–31] However, inherent geometric variation in such electrodes requires particular care during data analysis. This is necessary to disentangle the signal of the material under study from the direct tunneling current contributions, potential localized lead states, and to exclude reconnected graphene electrodes or connected graphene islands.[32–35] Moreover, the long fabrication time of each gap impacts the scalability of this approach.

Here, we report on graphene electrodes fabricated by electron beam lithography (EBL) using the combination of an optimized etch mask and etching recipe, which results in electrode separations down to <15 nm. This clean and well-defined electrode geometry helps to overcome the challenges emphasized above and represents an appealing platform to contact GNRs with a length of length above 15 nm. Moreover, the availability of large-scale graphene produced by chemical vapor deposition (CVD) allows us to fabricate up to 1680 devices per chip.

We demonstrate the suitability of our nanofabrication approach by integrating atomically precise 9-AGNRs in a field-effect transistor (FET) device.[36] 9-AGNRs are the ideal testing material due to their long-term stability and their well-studied transport properties.[37, 38] In addition, we show that thermal annealing is an efficient way to enhance the electrical device properties leading to an increase in the on-state current of up to an order of magnitude at room temperature. Moreover, our gate-dependent electrical transport measurements show on-off ratios reaching values as high as $10^4$. The results obtained in this work open perspectives for the integration of different types of GNRs in more complex device geometries.

## 2 Experimental

### 2.1 Graphene growth and transfer

Polycrystalline graphene is synthesized via chemical vapor deposition (CVD) in a tube furnace (Three zone HZS, Carbolite). A 25 μm thick copper (Cu) foil (Foil 2017, No. 46365, Alfa Aesar) is prepared at room temperature by first cleaning in acetone (15 min), rinsing in isopropanol (IPA), immersion in de-ionized (DI) water (5 min), acetic acid (30 min), DI-water (20 min + 5 min in an ultrasonic bath), ethanol (1 min) and blown dry with N2 before a reduction annealing in an H2-rich atmosphere (20 sccm H2 in 200 sccm Ar) at 1000 °C and <1 mbar for 60 min. Before the growth, the pressure inside the tube is increased to 110 mbar by partially closing the downstream valve. Graphene growth is initiated by the addition of 0.04 sccm CH4 to the chamber for 22 min. The growth is terminated by stopping the CH4 flow, reducing the pressure by opening the downstream valve, opening the lid of the tube furnace, and circulating air with a fan to allow for an abrupt drop in temperature. The cool down procedure (to <100 °C) takes 45 min. The as-grown single-layer graphene is transferred onto a specially-developed substrate for device integration of GNRs (target substrate) using a wet transfer method.[39] Poly(methyl methacrylate) (PMMA) 50K (AR-P 632.12, Allresist GmbH) is spun onto the graphene-coated Cu foil and the backside graphene is etched using reactive ion etching (RIE). Cu is etched away using a copper etchant (PC COPPER ETCHANT-100, Transene) for 60 min leaving the graphene/PMMA film floating on top. The etchant is then replaced by DI-water in a stepwise dilution process. The DI-water is then replaced by a 10 % HCl solution for 5min. After a final rinsing in DI-water, the graphene/PMMA film is fished out with the target substrate. After settling under ambient conditions for 30 min, the target substrate/graphene/PMMA stack is placed in an oven and heated to 80 °C for 1 h followed by a second heating step at 80 °C for at least 12 h under vacuum conditions (<1 mbar) to ensure good adhesion of the graphene to the target substrate. PMMA is removed in acetone for 10 min at room temperature, 60 min at 56 °C, followed by a 30 min cool down period. Finally, we perform an IPA rinsing step followed by N2 dry blowing. This process yields clean, single-layer graphene with low defect density on the target substrate, see Fig. S1.[40]

## 2.2 Graphene patterning

To define graphene electrodes, graphene on Si/SiO$_2$ with predefined metal electrodes and optimized areas for Raman spectroscopy is patterned by EBL as detailed below. Two exposure steps (100 kV write mode, EBPG5200, Raith GmbH) are done, each followed by an RIE step.[39]. For the first exposure step, the sample with graphene is spin-coated with 160 nm thick PMMA 50K (AR-P 632.06, Allresist GmbH) and 90 nm thick PMMA 950K (AR-P 672.02, Allresist GmbH), each baked at 180 °C on a hotplate for 5 min. Following a first electron beam exposure, the resist is developed in Methyl-isobutyl-ketone (MIBK):IPA (1:3) at room temperature for 60 seconds. RIE (15 sccm Ar, 30 sccm O$_2$, 25 W, 18 mTorr) for 30 s is used to remove the accessible graphene. PMMA is removed using acetone, IPA, and N$_2$ dry blow. After this pre-patterning of the graphene, a second EBL and RIE step (same etching plasma parameters as above, reduced time of 6 s) are carried out to separate the graphene electrodes.

Two different approaches for the fabrication of etch masks were investigated:

i) **CSAR mask**: In the first approach, a 60 nm thick CSAR resist (AR-P 6200.04, Allresist GmbH) is spin-coated. Following the second electron beam exposure, the resist is developed using a suitable developer (AR 600-546, Allresist GmbH) at room temperature for 1 min followed by an IPA rinse. After RIE, the etching mask is removed by immersing in 1-Methyl-2-pyrrolidinone (NMP) (Sigma Aldrich) at room temperature for 10 min followed by 60 min at 80 °C, cooled down for 30 min, rinsed with IPA, and blown dry with N$_2$.

ii) **PMMA mask and cold development:** In the second approach, a 60 nm thick layer of PMMA 950K (AR-P 672.02, Allresist GmbH) diluted in anisole (1:1) is spin-coated. The development of the resist after electron beam exposure is done in MIBK:IPA (1:3) at 2 °C for 45 s followed by an IPA rinse at 2 °C for 10 s. After RIE the etch mask is removed in the same way as after the first RIE step.

Both approaches yield clean and well-separated graphene electrodes with reproducible gap sizes (see Results and Supporting Information).

## 2.3  Graphene electrode separation

The separation of graphene electrodes (gap size) is assessed using scanning electron microscopy (SEM) (Helios 450, FEI) and atomic force microscopy (AFM) (Icon, Bruker) is employed to independently determine the electrode separation. The AFM is equipped with a sharp cantilever (tip radius = 2 nm) (SSS-NCHR-20, Nanosensors) operated in soft-tapping mode. The electrode separation by AFM is determined via a Python script, based on the nanoscope library.[41] Each line scan is smoothed individually using a Savitzy-Golay filter and the edges of the gap are determined by selecting the local maxima and minima in the first derivative on either side of the gap minimum. It was not possible to apply the same procedure to the SEM data due to the low contrast between the graphene and the $SiO_2$ of the target substrate and the small separation of the graphene electrodes. Therefore, the average and standard deviation are obtained from 20 manual measurements that are equally spaced along the gap.

## 2.4  Graphene quality after patterning

The patterned graphene electrodes are analyzed in air by 2D Raman mapping (Alpha300R, WITec) using a 488 nm incident laser beam at 1.5 mW and a 100x objective (NA = 0.9) with a pixel spacing of 100 nm.[39]

## 2.5  On-surface synthesis of 9-AGNRs and transfer to a device substrate

9-AGNRs were synthesized from 3′,6′-diiodo-1,1′:2′,1″-terphenyl (DITP).[10] Two types of GNRs, uniaxially aligned and randomly oriented, are transferred onto the sample. Using Au(788) as growth substrate results in uniaxially aligned 9-AGNRs (GNRs grown along the narrow (111) terraces) while using Au(111)/mica leads to non-aligned 9-AGNRs. [42, 43] In both cases, Au(788) single crystal (MaTeK,

Germany) or Au(111)/mica growth substrates (Phasis, Switzerland), are cleaned in ultrahigh vacuum by two sputtering/annealing cycles: 1 kV Ar$^+$ for 10 min followed by annealing at 420 °C for Au(788) and 470 °C for Au(111)/mica for 10 min. Next, the precursor monomer DITP is sublimed onto the Au surface from a quartz crucible heated to 70 °C, with the growth substrate held at room temperature. After deposition of 1 monolayer DITP, the growth substrate is heated (0.5 K/s) to 200 °C with a 10 min holding time to activate the polymerization reaction, followed by annealing at 400 °C (0.5 K/s with a 10 min holding time) to form the GNRs via cyclodehydrogenation.

The average GNR length is between 40 and 45 nm.[10] 9-AGNRs are transferred from their growth substrate to the silicon-based target substrates with predefined graphene electrodes by two different transfer approaches. 9-AGNRs grown on Au(788) crystals are transferred by an electrochemical delamination method using PMMA as described previously.[39, 44, 45] 9-AGNRs grown on Au(111)/mica are transferred using a polymer-free method as described elsewhere.[16, 37, 46]

## 2.6 Electronic measurements

All electronic measurements are performed under vacuum conditions (<10$^{-6}$ mbar) in two different probe stations. The FETs consisting of aligned GNRs using the polymer-assisted transfer are characterized in a custom-built probe station equipped with nanoprobes (miBot, Imina Technologies SA). A data acquisition board (USB-6289, National Instruments) is employed to apply the bias and gate voltages and read the voltage output of a custom-made I–V converter (Model SP983, Basel Precision Instruments GmbH). The FETs consisting of GNRs using the polymer-free transfer method are characterized in a commercially available probe station (Lake Shore Cryogenics, Model CRX-6.5K). A data acquisition board (ADwin-Gold II, Jäger Computergesteuerte Messtechnik GmbH) is employed to apply the bias and gate voltages and read the voltage output of the I–V converter (DDPCA-300, FEMTO Messtechnik GmbH).

# 3 Results and discussion

## 3.1 Graphene electrodes

The fabrication process yielding graphene electrodes for contacting graphene nanoribbons is illustrated in Fig. 1 and described in detail in the experimental section.

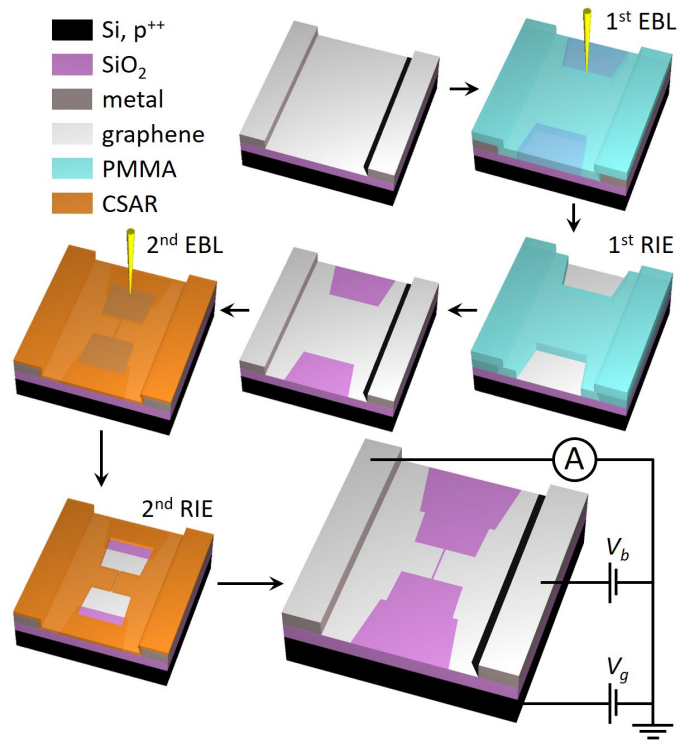

**FIG. 1.** Fabrication procedure and electrical measurement configuration. Three-dimensional illustration describing the fabrication steps of the EBL defined graphene electrodes. The resist regions exposed during electron beam lithography are marked with darker colors. Arrows indicate fabrication order. The electrical measurement configuration is schematically shown in the last, slightly larger, illustration. See main text for details.

We stress that for etching nanogaps into graphene the interplay of the used etch masks, their removal, as well as the chosen etching parameters, plays an even more crucial role in obtaining the wanted feature resolution than for evaporated features.

First, it is crucial to split the resist exposure by electron beam into two steps to ensure the proximity-effect while writing the coarse features does not affect the sensitive exposure of the nanogap. For the second exposure, we use a beam step size of 5 nm and beam current of 3 nA resulting in a beam diameter of 5 nm. Second, for the first RIE step to pattern the coarse features, a double layer etch mask is employed that helps reduce contaminations on graphene by using a low molecular weight resist in direct contact with the graphene and a high molecular weight resist on top for high contrast and feature definition. We note that the undercut in the double layer resist does not pose a problem, as the feature sizes in this first RIE step are not critical. For the second RIE step, an undercut is unwanted since it would result in a larger electrode separation. The smallest graphene electrode separation (<15 nm) is achieved using CSAR resist, due to its excellent performance in terms of resolution, sensitivity, and etch resistance.[47]

Third, we emphasize that the duration of the second RIE step has to be short enough to avoid a sideways etching of the resist mask but long enough that the monolayer graphene is fully etched. This trade-off leads to a delicate balance between device yield and electrode separation. Last, the removal of the etch mask has to be done using processes that are sufficiently mild to preserve graphene's quality but sufficiently harsh to leave little residues on the electrodes. We therefore, employed only acetone and NMP since their effects on graphene's quality are well studied.[48–51]

Since the used GNR growth substrates do not exceed 5x5 mm in size, after fabrication the chip is broken into smaller pieces with 100-200 devices each before the GNRs to target substrate transfer.

## 3.2 Characterization of patterned graphene electrodes

Before the electrode separation is assessed, an optical inspection of the graphene electrodes is carried out. Graphene electrodes containing graphene folds in the central region or those damaged during the fabrication process are excluded from further investigation. A typical optical image of the graphene electrodes can be seen in Fig. 2(a).

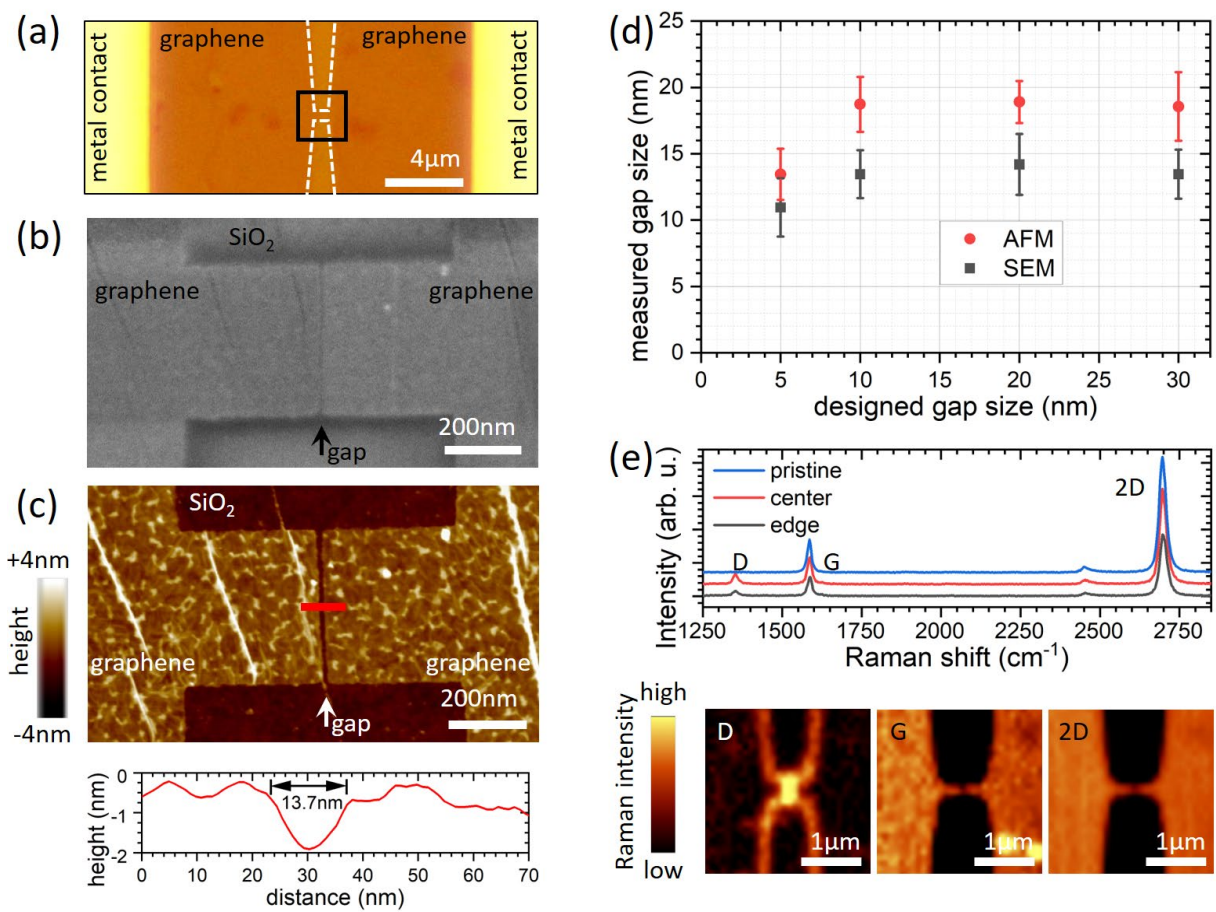

**FIG. 2.** Electrode separation of devices fabricated using CSAR mask. (a) Optical image of a representative device. Of the central region of the same device is shown in (b) a SEM image and in (c) a height profile (AFM scan) including a line cut through the central region (red) showing the electrode separation. (d) Electrode separations measured by SEM and AFM for four devices with different gap sizes. (e) Raman intensity maps for D-, G- and 2D-bands in the area indicated with a black square in (a) and spectra extracted at representative positions.

The electrode separation is assessed by SEM and AFM and representative scans are shown in Fig. 2(b) and 2(c), respectively, with in Fig. 2(d) extracted gap sizes. We find that the graphene electrodes fabricated with the CSAR etch mask (see experimental section) are separated by <15 nm for the smallest designed geometry. The fabrication method using the PMMA etch mask yields a slightly larger electrode separation of around 27 nm in the smallest case (See Fig. S3). Hence, these electrodes are only used for the uniaxially aligned 9-AGNRs in order to have high device yields. Fig. 2 (d) also shows that the measured electrode separation for the CSAR etch mask does not scale linearly with the width of the gap in the design. We attribute this behavior to the proximity-effect correction procedure that is applied for the exposure dose calculation.

For a successful device integration of GNRs, we consider it important to have graphene electrodes with little to no defects after patterning. Raman spectroscopy maps confirm the high quality of the graphene electrodes after processing (see Fig. 2(e)). The D-band intensity map shows in the pristine area neglectable intensity and an intensity increase at the edges and in the nanogap region. Raman spectroscopy further revealed a clear drop in the intensity of the G- and 2D- bands where the graphene electrodes are separated, indicating a lower amount of carbon and breaking of the crystal structure. The 488 nm excitation source was chosen to reach a minimal laser spot size for best spatial resolution. The high graphene quality was also confirmed by measuring the current versus applied gate voltage of a reference device that underwent the same fabrication procedure except for the 2$^{nd}$ RIE step, revealing field-effect mobilities of 2'500 cm$^2$/Vs for electrons and 1'800 cm$^2$/Vs for holes (see Fig. S3).

We optically assessed 91 devices for the two transfer methods based on which we excluded 24 devices. After the initial optical assessment and prior to the 9-AGNR transfer, each device is characterized electrically. A schematic illustration of the electronic wiring for the latter is depicted in Fig. 3(a). As shown in Fig. 3(b) a high yield of clearly separated graphene electrodes (>1 T$\Omega$) of 79.1 % is found. The remaining 20.9 % of graphene electrodes are either weakly (<1 T$\Omega$) or fully connected (<1 G$\Omega$). Representative I-V characteristics for the three cases are shown in Fig. 3(c).

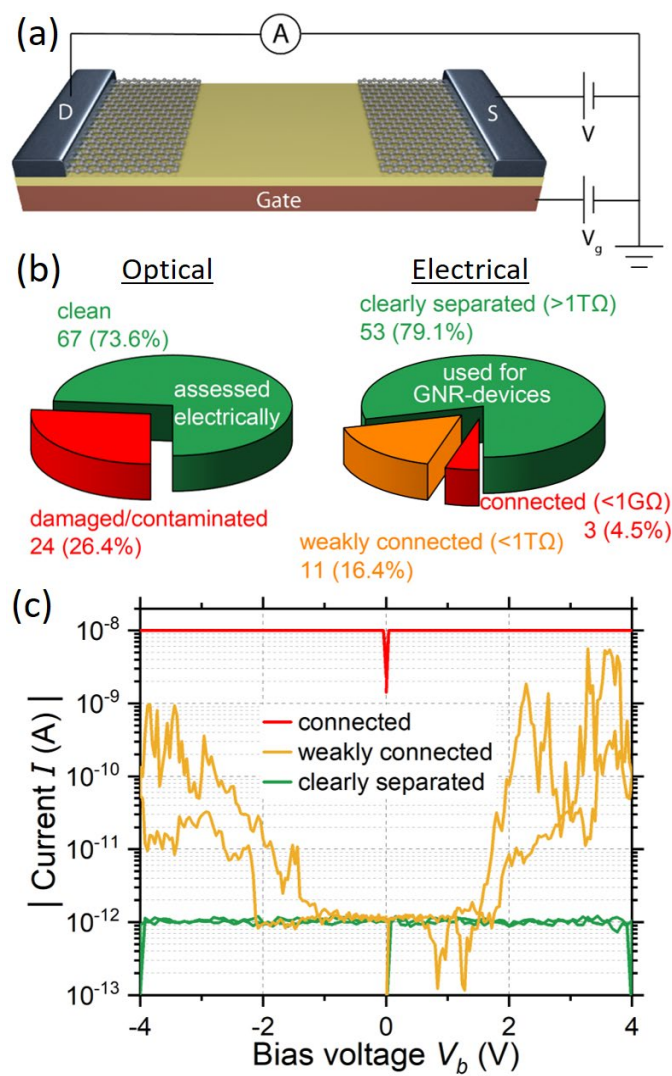

**FIG. 3**. (a) Illustration of the electrical measurement scheme for characterizing the separation of the graphene electrodes. (b) Statistics on electrode separation by optical and electrical assessment. (c) I-V characteristics of three representative devices before 9-AGNR transfer.

Reasons for the not well formed nanogaps may include the presence of (partially etched) multilayer graphene at the constriction and contamination during/after processing. These devices are not investigated further.

## 3.3 Electrical characterization and effect of annealing

After the initial characterization of the devices, 9-AGNRs were transferred on top of the graphene electrodes. Fig. S4 shows high resolution scanning tunneling microscope (STM) images of 9-AGNRs on the growth substrates presenting their alingment. We note that the growth and transfer method were adapted to the gapsize. For the devices fabricated using the PMMA mask (gapsize ~28 nm), uniaxially aligned GNRs were transferred oriented perpendicular to the gap to maximize the chance of bridging both electrodes. For the devices fabricated using the CSAR mask (gapsize <15 nm), based on geometrical considerations, we anticipate a higher probability of bridging that allows for investigating non-aligned 9-AGNRs transferred using a polymer-free method. In both cases, the integrity of the 9-AGNRs after the transfer process was confirmed by Raman spectroscopy (Fig. S5). In particular, the presence of the longitudinal compressive mode (LCM) is strong evidence for the high quality of the 9-AGNRs after the transfer process due to its high sensitivity to structural damage.[45] Fig. 4(a) shows typical I-V characteristics recorded at room temperature under vacuum conditions (<$10^{-6}$ mbar) on different devices fabricated using the two transfer methods. We observe highly nonlinear I-V curves with currents up to 0.5 nA at 1 V bias voltage. The inset presents a schematic of the device and the electrical characterization scheme.

To improve the maximum currents through the devices, we investigated the effect of thermal annealing. As the transfer of the 9-AGNRs onto the target substrate exposes the graphene to humidity and even water in the case of the polymer assisted transfer, the samples were heated to 150 °C for 30 minutes at $10^{-6}$ mbar to remove water residues at the graphene/GNR interface. The heating also provides energy for local geometric rearrangements. To evaluate the benefit of this thermal treatment, the maximum currents observed at a gate voltage of 0 V and a bias voltage of 1 V are compared before (as transferred, $I_{transferred}$) and after thermal annealing ($I_{annealed}$). Fig. 4(b) shows a scatter plot of the ratio $I_{annealed} / I_{transferred}$ for all devices. Over all samples, an increase by one order of magnitude or higher in 50 % of the devices is observed, with individual junctions showing an increase as high as a factor of 100.

During annealing, several processes can take place and affect the conductance of the junctions. By reducing the number of water adsorbates at the GNR-graphene interface, the two nanomaterials can go into a more intimate contact, which can lead to an increased electronic coupling similar to what has been observed for decoupled graphene monolayers.[52, 53] Water removal may also result in reduced doping of the GNRs, leading to a probing of the more intrinsic GNR transport properties, similar to what has been reported for graphene FETs on $SiO_2$.[54, 55] Studies of molecules with planar anchor groups on graphene electrodes revealed that the binding energy to sliding and bending is around 0.01 eV, significantly lower than the energy $k_BT$ (~0.04 eV) provided during the thermal annealing process.[56–58] Hence, this energy likely is sufficient to cause local displacement and geometrical rearrangements of the GNRs that can lead to both improved contact but also loss of GNRs within the junction and a decrease of overall conductance in multi-GNR junctions as can be seen by a conductance decrease in about 20 % of the devices after annealing. We note that at the annealing temperature used, no lateral fusion of GNRs is expected.[30, 59]

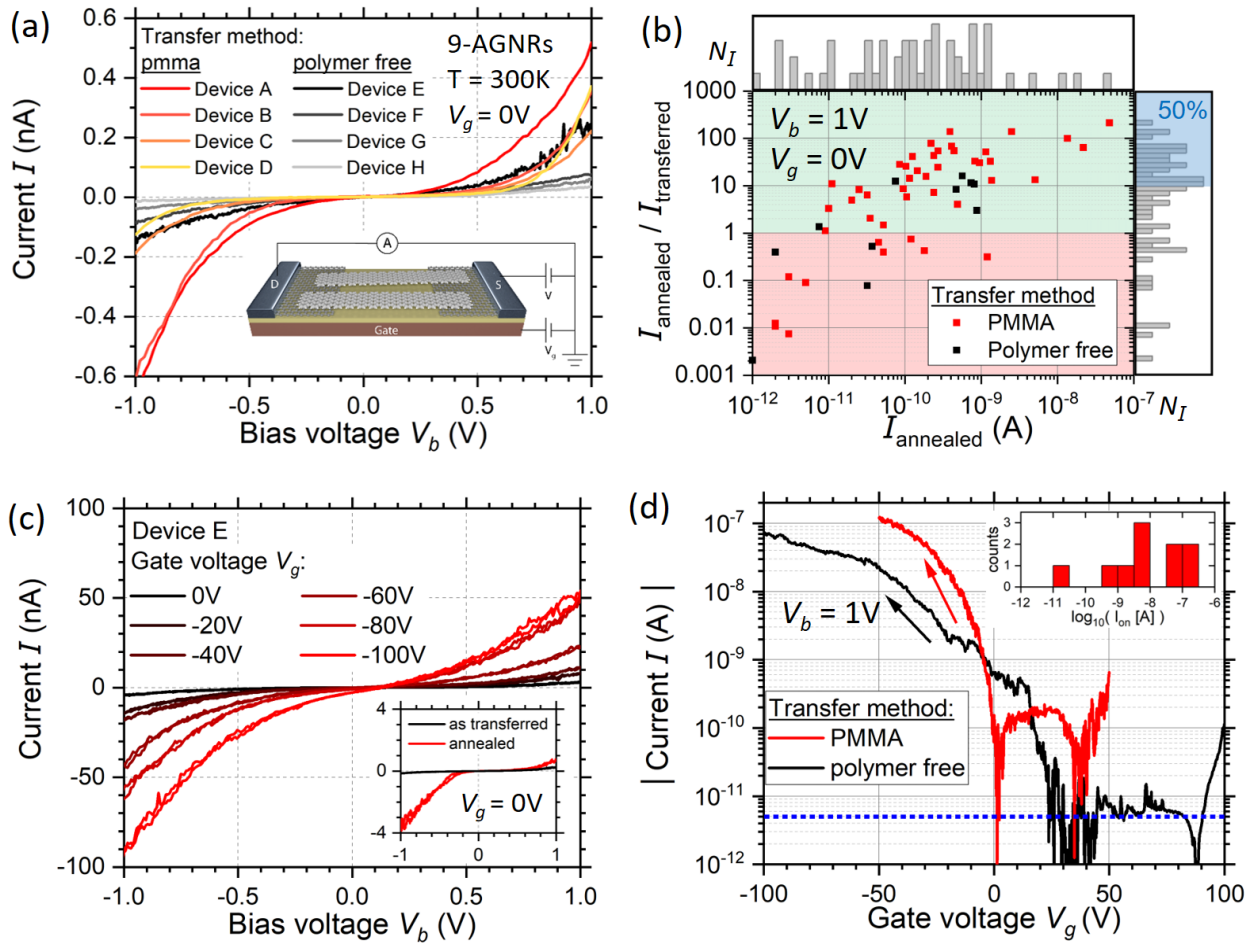

**FIG. 4.** (a) I-V characteristics of four representative 9-AGNR devices for each of the two transfer methods before thermal annealing. (b) The main panel shows the ratio of current values measured before and after the annealing of the devices. Current increase and decrease are highlighted in green and red respectively. Histograms of $I_{transferred}$ and $I_{annealed} / I_{transferred}$ are shown in the outer panels on the top and the right respectively. Highlighted in blue are 50 % of the devices. (c) I-V characteristics of device

E recorded at different applied gate voltages after annealing. Inset shows the effect of annealing on the I-V characteristics. (d) Current – gate voltage dependence at fixed bias voltage for two devices. Arrows indicate the direction of the gate sweep. The blue dashed line indicates the noise floor of the I-V converter. The inset shows a histogram of the on-currents.

In Fig. 4(c), we measured current-voltage characteristics at various applied gate voltages. The plot indicates slightly asymmetric characteristics with a strong gate dependence. The presence of only little hysteresis effects between the up and down sweep of the applied bias voltage indicates high device stability. The observed hysteresis is attributed to the influence of trap states in the oxide.[60] Fig. 4(d) shows a measurement of the current as a function of gate voltage for two devices, recorded at a fixed applied bias voltage of 1 V (extended data see Fig. S6). The traces show a drastic increase of the conductance for negative gate voltages, pointing towards hole transport through the valence band (or highest occupied molecular orbital (HOMO)).[12] The gentle increase in conductance at positive gate voltages suggests the presence of another transport channel entering the bias window, presumably the conduction band (or lowest unoccupied molecular orbital (LUMO)). In the gate sweeps, we obtain maximal on-off ratios of up to $10^4$, with an on-current of 70 nA at a gate voltage of -100 V and an off-current of 5 pA at a gate voltage of +40 V for the black curve in Fig. 4(d). The maximum observed values for the on-off ratios is about a factor of 100 higher than reported by Martini *et al.* and El Abbassi *et al.* contacting 9-AGNRs using graphene electrodes fabricated by electrical breakdown, and about a factor 50 higher than reported by Jangid *et al.* for top-down fabricated GNRs.[11, 30, 61]

## 4  Conclusions

We successfully integrated 9-AGNRs in a FET geometry using graphene electrodes fabricated via optimized e-beam lithography and reactive ion etching resulting in electrode separations as small as <15 nm. Room-temperature electrical transport measurements revealed nonlinear current-voltage characteristics and a strong gate dependence. Furthermore, we find that thermal annealing improves the on-currents after annealing by at least one order of magnitude in 50% of the investigated devices. In addition, we performed gate sweeps revealing on-off ratios as high as $10^4$ with the highest on-currents of 70 nA at a bias voltage of 1 V.

The developed technology to fabricate graphene electrodes separated by <15 nm is a major step forward towards all-carbon electronics and offers encouraging prospects for room-temperature ambipolar 9-AGNR-FET behavior. The presented platform could also be applied for short channel FETs using two-dimensional materials as channel material, as reported for $MoS_2$ or phase-change memory devices.[62, 63] Importantly, this platform will allow for the integration of GNRs of different widths as well as different edge structures for exploring more exotic transport properties.[13, 64]

CRediT authorship contribution statement

**Oliver Braun:** Conceptualization, Methodology, Investigation, Writing – Original Draft, Writing – Review & Editing, Visualization, Project administration. **Jan Overbeck:** Conceptualization, Methodology, Investigation, Writing – Original Draft, Visualization. **Maria El Abbassi:** Investigation. **Silvan Käser:** Investigation. **Roman Furrer:** Resources. **Antonis Olziersky:** Investigation. **Alexander Flasby:** Software. **Gabriela Borin Barin:** Writing – Original Draft, Resources. **Rimah Darawish:** Resources. **Klaus Müllen:** Resources. **Pascal Ruffieux:** Supervision. **Roman Fasel:** Funding acquisition, Supervision. **Ivan Shorubalko:** Conceptualization, Methodology. **Mickael L. Perrin:** Writing – Original Draft, Writing – Review & Editing, Visualization, Supervision. **Michel Calame: Conceptualization,** Writing – Original Draft, Writing – Review & Editing, Visualization, Funding acquisition, Supervision

Declaration of interests

The authors declare that they have no known competing financial interests or personal relationships that could have appeared to influence the work reported in this paper.

Acknowledgments

We are grateful for the support provided by the Binnig and Rohrer Nanotechnology Center (BRNC). O.B. thanks Michael Stiefel for the graphene electrode imaging using SEM. This work was in part supported by the European Union under the FET open project QuIET no. 767187. M.L.P. acknowledges funding by the EMPAPOSTDOCS-II program which is financed by the European Union Horizon 2020 research and innovation program under the Marie Skłodowska-Curie grant agreement number 754364 and the Swiss National Science Foundation (SNSF) under the Spark project no. 196795. G.B.B, R.D, P.R and R.F. acknowledge funding by the Swiss National Science Foundation under grant no. 200020_182015, the European Union Horizon 2020 research and innovation program under grant agreement no. 881603 (GrapheneFlagship Core 3), and the Office of Naval Research BRC Program under the grant N00014-18-1-2708. Finally, we acknowledge the Scanning Probe Microscopy User lab at Empa for providing access to the AFM setup.

Data Availability

The data that support the findings of this study are available from the corresponding author upon reasonable request.

Appendix A. Supplementary information

Supplementary information to this article is available free of charge.

# References

[1]  Iijima S. Helical microtubules of graphitic carbon. Nature 1991;354(6348):56–8.

[2]  Tans SJ, Devoret MH, Dai H, Thess A, Smalley RE, Geerligs LJ et al. Individual single-wall carbon nanotubes as quantum wires. Nature 1997;386(6624):474–7.

[3]  Mann D, Javey A, Kong J, Wang Q, Dai H. Ballistic Transport in Metallic Nanotubes with Reliable Pd Ohmic Contacts. Nano Lett. 2003;3(11):1541–4.

[4]  Samm J, Gramich J, Baumgartner A, Weiss M, Schönenberger C. Optimized fabrication and characterization of carbon nanotube spin valves. Journal of Applied Physics 2014;115(17):174309.

[5]  Hills G, Lau C, Wright A, Fuller S, Bishop MD, Srimani T et al. Modern microprocessor built from complementary carbon nanotube transistors. Nature 2019;572(7771):595--602.

[6]  Treuting RG, Arnold SM. Orientation habits of metal whiskers. Acta Metallurgica 1957;5(10):598.

[7]  Gangadharaiah S, Braunecker B, Simon P, Loss D. Majorana edge states in interacting one-dimensional systems. Physical review letters 2011;107(3):36801.

[8]  Kloeffel C, Loss D. Prospects for Spin-Based Quantum Computing in Quantum Dots. Annu. Rev. Condens. Matter Phys. 2013;4(1):51–81.

[9]  Cui Y, Zhong Z, Wang D, Wang WU, Lieber CM. High Performance Silicon Nanowire Field Effect Transistors. Nano Lett. 2003;3(2):149–52.

[10]  Di Giovannantonio M, Deniz O, Urgel JI, Widmer R, Dienel T, Stolz S et al. On-Surface Growth Dynamics of Graphene Nanoribbons: The Role of Halogen Functionalization. ACS nano 2018;12(1):74–81.

[11]  El Abbassi M, Perrin ML, Barin GB, Sangtarash S, Overbeck J, Braun O et al. Controlled Quantum Dot Formation in Atomically Engineered Graphene Nanoribbon Field-Effect Transistors. ACS nano 2020;14(5):5754–62.

[12]  Llinas JP, Fairbrother A, Borin Barin G, Shi W, Lee K, Wu S et al. Short-channel field-effect transistors with 9-atom and 13-atom wide graphene nanoribbons. Nature communications 2017;8(1):633.

[13]  Sun Q, Gröning O, Overbeck J, Braun O, Perrin ML, Borin Barin G et al. Massive Dirac Fermion Behavior in a Low Bandgap Graphene Nanoribbon Near a Topological Phase Boundary. Adv. Mater. 2020;32(12):1906054.

[14]  Yamaguchi J, Hayashi H, Jippo H, Shiotari A, Ohtomo M, Sakakura M et al. Small bandgap in atomically precise 17-atom-wide armchair-edged graphene nanoribbons. Commun Mater 2020;1(1):17954.

[15]  Bennett PB, Pedramrazi Z, Madani A, Chen Y-C, Oteyza DG de, Chen C et al. Bottom-up graphene nanoribbon field-effect transistors. Appl. Phys. Lett. 2013;103(25):253114.

[16] Fairbrother A, Sanchez-Valencia J-R, Lauber B, Shorubalko I, Ruffieux P, Hintermann T et al. High vacuum synthesis and ambient stability of bottom-up graphene nanoribbons. Nanoscale 2017;9(8):2785–92.

[17] Chen Z, Zhang W, Palma C-A, Lodi Rizzini A, Liu B, Abbas A et al. Synthesis of Graphene Nanoribbons by Ambient-Pressure Chemical Vapor Deposition and Device Integration. Journal of the American Chemical Society 2016;138(47):15488–96.

[18] Ruffieux P, Wang S, Yang B, Sánchez-Sánchez C, Liu J, Dienel T et al. On-surface synthesis of graphene nanoribbons with zigzag edge topology. Nature 2016;531(7595):489–92.

[19] Rizzo DJ, Veber G, Cao T, Bronner C, Chen T, Zhao F et al. Topological band engineering of graphene nanoribbons. Nature 2018;560(7717):204–8.

[20] Gröning O, Wang S, Yao X, Pignedoli CA, Borin Barin G, Daniels C et al. Engineering of robust topological quantum phases in graphene nanoribbons. Nature 2018;560(7717):209–13.

[21] Keerthi A, Radha B, Rizzo D, Lu H, Diez Cabanes V, Hou IC-Y et al. Edge Functionalization of Structurally Defined Graphene Nanoribbons for Modulating the Self-Assembled Structures. Journal of the American Chemical Society 2017;139(46):16454–7.

[22] Richter N, Chen Z, Tries A, Prechtl T, Narita A, Müllen K et al. Charge transport mechanism in networks of armchair graphene nanoribbons. Scientific reports 2020;10(1):1988.

[23] Perrin ML, Verzijl CJO, Martin CA, Shaikh AJ, Eelkema R, van Esch JH et al. Large tunable image-charge effects in single-molecule junctions. Nature nanotechnology 2013;8(4):282–7.

[24] Perrin ML, Burzurí E, van der Zant HSJ. Single-molecule transistors. Chemical Society reviews 2015;44(4):902–19.

[25] Moth-Poulsen K, Bjørnholm T. Molecular electronics with single molecules in solid-state devices. Nature nanotechnology 2009;4(9):551–6.

[26] Yu WJ, Liu Y, Zhou H, Yin A, Li Z, Huang Y et al. Highly efficient gate-tunable photocurrent generation in vertical heterostructures of layered materials. Nature nanotechnology 2013;8(12):952–8.

[27] Liu Y, Wu H, Cheng H-C, Yang S, Zhu E, He Q et al. Toward barrier free contact to molybdenum disulfide using graphene electrodes. Nano letters 2015;15(5):3030–4.

[28] El Abbassi M, Pósa L, Makk P, Nef C, Thodkar K, Halbritter A et al. From electroburning to sublimation: substrate and environmental effects in the electrical breakdown process of monolayer graphene. Nanoscale 2017;9(44):17312–7.

[29] El Abbassi M, Sangtarash S, Liu X, Perrin ML, Braun O, Lambert C et al. Robust graphene-based molecular devices. Nature nanotechnology 2019;14(10):957–61.

[30] Martini L, Chen Z, Mishra N, Barin GB, Fantuzzi P, Ruffieux P et al. Structure-dependent electrical properties of graphene nanoribbon devices with graphene electrodes. Carbon 2019;146:36–43.

[31] Candini A, Martini L, Chen Z, Mishra N, Convertino D, Coletti C et al. High Photoresponsivity in Graphene Nanoribbon Field-Effect Transistor Devices Contacted with Graphene Electrodes. J. Phys. Chem. C 2017;121(19):10620–5.

[32] Gehring P, Sadeghi H, Sangtarash S, Lau CS, Liu J, Ardavan A et al. Quantum Interference in Graphene Nanoconstrictions. Nano letters 2016;16(7):4210–6.

[33] García-Suárez VM, García-Fuente A, Carrascal DJ, Burzurí E, Koole M, van der Zant HSJ et al. Spin signatures in the electrical response of graphene nanogaps. Nanoscale 2018;10(38):18169–77.

[34] Gehring P, Sowa JK, Cremers J, Wu Q, Sadeghi H, Sheng Y et al. Distinguishing Lead and Molecule States in Graphene-Based Single-Electron Transistors. ACS nano 2017;11(6):5325–31.

[35] Barreiro A, van der Zant HSJ, Vandersypen LMK. Quantum dots at room temperature carved out from few-layer graphene. Nano letters 2012;12(12):6096–100.

[36] Talirz L, Söde H, Dumslaff T, Wang S, Sanchez-Valencia JR, Liu J et al. On-Surface Synthesis and Characterization of 9-Atom Wide Armchair Graphene Nanoribbons. ACS nano 2017;11(2):1380–8.

[37] Borin Barin G, Fairbrother A, Rotach L, Bayle M, Paillet M, Liang L et al. Surface-Synthesized Graphene Nanoribbons for Room Temperature Switching Devices: Substrate Transfer and ex Situ Characterization. ACS Appl. Nano Mater. 2019;2(4):2184–92.

[38] Pizzochero M, Borin Barin G, Ruffieux P, Fasel R. Quantum Electronic Transport Across "Bite" Defects in Graphene Nanoribbons. arXiv preprint 2020:arXiv:2006.15075.

[39] Overbeck J, Borin Barin G, Daniels C, Perrin ML, Liang L, Braun O et al. Optimized Substrates and Measurement Approaches for Raman Spectroscopy of Graphene Nanoribbons. Phys. Status Solidi B 2019;256(12):1900343.

[40] Hong J-Y, Shin YC, Zubair A, Mao Y, Palacios T, Dresselhaus MS et al. A Rational Strategy for Graphene Transfer on Substrates with Rough Features. Advanced materials (Deerfield Beach, Fla.) 2016:2382–92.

[41] jmarini. nanoscope 0.12.1, https://pypi.org/project/nanoscope/; 2017.

[42] Linden S, Zhong D, Timmer A, Aghdassi N, Franke JH, Zhang H et al. Electronic structure of spatially aligned graphene nanoribbons on Au(788). Physical review letters 2012;108(21):216801.

[43] Ruffieux P, Cai J, Plumb NC, Patthey L, Prezzi D, Ferretti A et al. Electronic structure of atomically precise graphene nanoribbons. ACS nano 2012;6(8):6930–5.

[44] Senkovskiy BV, Pfeiffer M, Alavi SK, Bliesener A, Zhu J, Michel S et al. Making Graphene Nanoribbons Photoluminescent. Nano letters 2017;17(7):4029–37.

[45] Overbeck J, Barin GB, Daniels C, Perrin ML, Braun O, Sun Q et al. A Universal Length-Dependent Vibrational Mode in Graphene Nanoribbons. ACS nano 2019;13(11):13083–91.

[46] Backes C, Abdelkader AM, Alonso C, Andrieux-Ledier A, Arenal R, Azpeitia J et al. Production and processing of graphene and related materials. 2D Mater. 2020;7(2):22001.

[47] Thoms S, Macintyre DS. Investigation of CSAR 62, a new resist for electron beam lithography. Journal of Vacuum Science & Technology B, Nanotechnology and Microelectronics: Materials, Processing, Measurement, and Phenomena 2014;32(6):06FJ01.

[48] Chen Y, Gong X-L, Gai J-G. Progress and Challenges in Transfer of Large-Area Graphene Films. Advanced science (Weinheim, Baden-Wurttemberg, Germany) 2016;3(8):1500343.

[49] Chen P-C, Lin C-P, Hong C-J, Yang C-H, Lin Y-Y, Li M-Y et al. Effective N-methyl-2-pyrrolidone wet cleaning for fabricating high-performance monolayer MoS2 transistors. Nano Res. 2019;12(2):303–8.

[50] Zhuang B, Li S, Li S, Yin J. Ways to eliminate PMMA residues on graphene —— superclean graphene. Carbon 2021;173(5696):609–36.

[51] Thodkar K, Thompson D, Lüönd F, Moser L, Overney F, Marot L et al. Restoring the Electrical Properties of CVD Graphene via Physisorption of Molecular Adsorbates. ACS applied materials & interfaces 2017;9(29):25014–22.

[52] Schmidt H, Lüdtke T, Barthold P, McCann E, Fal'ko VI, Haug RJ. Tunable graphene system with two decoupled monolayers. Appl. Phys. Lett. 2008;93(17):172108.

[53] Rickhaus P, Liu M-H, Kurpas M, Kurzmann A, Lee Y, Overweg H et al. The electronic thickness of graphene. Science advances 2020;6(11):eaay8409.

[54] Cheng Z, Zhou Q, Wang C, Li Q, Wang C, Fang Y. Toward intrinsic graphene surfaces: a systematic study on thermal annealing and wet-chemical treatment of SiO2-supported graphene devices. Nano letters 2011;11(2):767–71.

[55] Hong Y, Wang S, Li Q, Song X, Wang Z, Zhang X et al. Interfacial icelike water local doping of graphene. Nanoscale 2019;11(41):19334–40.

[56] Bailey S, Visontai D, Lambert CJ, Bryce MR, Frampton H, Chappell D. A study of planar anchor groups for graphene-based single-molecule electronics. The Journal of chemical physics 2014;140(5):54708.

[57] Robinson BJ, Bailey SWD, O'Driscoll LJ, Visontai D, Welsh DJ, Mostert AB et al. Formation of Two-Dimensional Micelles on Graphene: Multi-Scale Theoretical and Experimental Study. ACS nano 2017;11(3):3404–12.

[58] Sadeghi H, Sangtarash S, Lambert C. Robust Molecular Anchoring to Graphene Electrodes. Nano letters 2017;17(8):4611–8.

[59] Chen Z, Wang HI, Bilbao N, Teyssandier J, Prechtl T, Cavani N et al. Lateral Fusion of Chemical Vapor Deposited N = 5 Armchair Graphene Nanoribbons. Journal of the American Chemical Society 2017;139(28):9483–6.

[60] Tries A, Richter N, Chen Z, Narita A, Müllen K, Wang HI et al. Hysteresis in graphene nanoribbon field-effect devices. Physical chemistry chemical physics PCCP 2020;22(10):5667–72.

[61] Jangid P, Pathan D, Kottantharayil A. Graphene nanoribbon transistors with high ION/IOFF ratio and mobility. Carbon 2018;132(7419):65–70.

[62] Xie L, Liao M, Wang S, Yu H, Du L, Tang J et al. Graphene-Contacted Ultrashort Channel Monolayer MoS2 Transistors. Advanced materials (Deerfield Beach, Fla.) 2017;29(37).

[63] Behnam A, Xiong F, Cappelli A, Wang NC, Carrion EA, Hong S et al. Nanoscale phase change memory with graphene ribbon electrodes. Appl. Phys. Lett. 2015;107(12):123508.

[64] Liu X, Li G, Lipatov A, Sun T, Mehdi Pour M, Aluru NR et al. Chevron-type graphene nanoribbons with a reduced energy band gap: Solution synthesis, scanning tunneling microscopy and electrical characterization. Nano Res. 2020;13(6):1713–22.